\documentclass{article}
\usepackage{spconf,amsmath,graphicx}
\usepackage{xcolor}
\usepackage{hyperref}

\title{FBWave: Efficient and Scalable Neural Vocoders for Streaming Text-To-Speech on the Edge}
\name{Bichen Wu$^1$, Qing He$^2$, Peizhao Zhang$^1$, Thilo Koehler$^2$, Kurt Keutzer$^3$, Peter Vajda$^1$\thanks{Correspondence to Bichen Wu: wbc@fb.com}}
\address{$^1$Facebook Reality Labs, $^2$Facebook AI, $^3$UC Berkeley}

\begin{document}
%
\maketitle
\begin{abstract}
Nowadays more and more applications can benefit from edge-based text-to-speech (TTS). However, most existing TTS models are too computationally expensive and are not flexible enough to be deployed on the diverse variety of edge devices with their equally diverse computational capacities. To address this, we propose FBWave, a family of efficient and scalable neural vocoders that can achieve optimal performance-efficiency trade-offs for different edge devices. FBWave is a hybrid flow-based generative model that combines the advantages of autoregressive and non-autoregressive models. It produces high quality audio and supports streaming during inference while remaining highly computationally efficient. Our experiments show that FBWave can achieve similar audio quality to WaveRNN while reducing MACs by 40x. More efficient variants of FBWave can achieve up to 109x fewer MACs while still delivering acceptable audio quality. Audio demos are available at \url{https://bichenwu09.github.io/vocoder_demos}. 
\end{abstract}
\begin{keywords}
Text-to-Speech, Generative Models, Normalizing Flows, Efficient Deep Learning, Edge Computing 
\end{keywords}
\section{Introduction}
\label{sec:intro}

Voice user interfaces (VUI) are absolutely essential for some applications and desirable for many others. A key component in these interfaces is text-to-speech (TTS), which translates text into speech audio. Previous TTS systems are mostly deployed on cloud servers with generated audios sent to users through the internet. However, in recent years, due to privacy concerns, concerns around internet stability and availability, and fast progress of edge computing, increasingly more applications would require or benefit from deploying TTS on devices such as smart watches, AR/VR glasses, home assistant devices, and mobile phones. 

This adds new challenges for TTS technologies in several ways: 1) Although previous TTS systems are able to generate speech audios with good quality \cite{wavenet,waveglow,WaveRNN}, their computational costs are prohibitively high for edge devices. 2) Deploying TTS to the edge entails dealing with a greater diversity of devices, with each device having its own computational budget. For example, a TTS model that can run on mobile phones may still be too big for smart watches. To cope with this, we not only need one efficient model, but a family of efficient and scalable models that can achieve good quality-efficiency trade-offs under different computational constraints. Our paper aims to address these challenges. 

Current state-of-the-art TTS models usually contain two components: the first one generates acoustic features from input text \cite{wang2017tacotron,tacotron2,MSTransformerTTS,ren2019fastspeech}; and the second model, also referred to as a vocoder, accepts the acoustic features and generates audio samples. Since a vocoder needs to generate samples at a high frequency, typically 16-24K samples per second, it is usually the computational bottleneck of a TTS system. Our paper is focused on vocoders. Despite the recent progress of vocoders, previous work has not been able to fully address the challenges of edge-based TTS. Early success of neural vocoders are mainly autoregressive models, such as WaveNet \cite{wavenet}, WaveRNN \cite{WaveRNN}. These models can generate high fidelity audios, but they are not efficient, due to their high model complexity and the nature of autoregression. Recent research efforts have been focusing on building non-autoregressive models, such as WaveGlow \cite{waveglow}. WaveGlow is a generative flow-based model that can synthesize high-fidelity audios using a parallel architecture. However, the model is extremely computationally expensive, as it requires 231 GMACs to compute 1 second of 22kHz audio, which is far beyond the capacity of edge devices. Following WaveGlow, SqueezeWave \cite{zhai2020squeezewave} shows that the computational cost of WaveGlow can be reduced by modifying the model to operate at a coarser temporal resolution. In addition, MelGAN and variants \cite{kumar2019melgan, yang2020multi} also adopt non-autoregressive architectures and train the models following the paradigm of GAN \cite{goodfellow2014generative}. However, audios generated by non-autoregressive models suffer from a periodic artifact since their output audio samples lacks continuity, and the quality further deteriorates when used in a streaming pipeline. 

To overcome this, we propose FBWave, a hybrid flow-based model that consists of a non-autoregressive convolution-based flow (ConvFlow) and an autoregressive GRU-based flow (GRUFlow). The GRUFlow mitigates the artifacts in WaveGlow and SqueezeWave by improving the continuity in generated samples. This allows us to use fewer layers of ConvFlows and let ConvFlows operate at a lower frequency, both of which lead to significant MAC reduction. It also enables the model to operate in a streaming pipeline. In ConvFlow and GRUFlow, we adopted inverted residual blocks to replace the complex WaveNet-like functions used in WaveGlow \cite{waveglow} and SqueezeWave \cite{zhai2020squeezewave}, which leads to MAC reduction and faster training. FBWave achieves much better audio quality than SqueezeWave \cite{zhai2020squeezewave}, but remains to be computationally efficient and requires much fewer MACs than previous work \cite{waveglow, WaveRNN}. To adapt to different devices with different budgets, we control the FBWave using a few architecture hyperparameters to obtain a family of models with different quality-efficiency trade-offs. Our experiments show that FBWave can achieve similar audio quality compared to WaveRNN \cite{WaveRNN} while using 40x fewer MACs and a smaller FBWave can deliver acceptable audio quality using 109x fewer MACs. 

\section{Method}
\label{sec:method}

\begin{figure}[h]
    \centering
    \includegraphics[width=.9\linewidth]{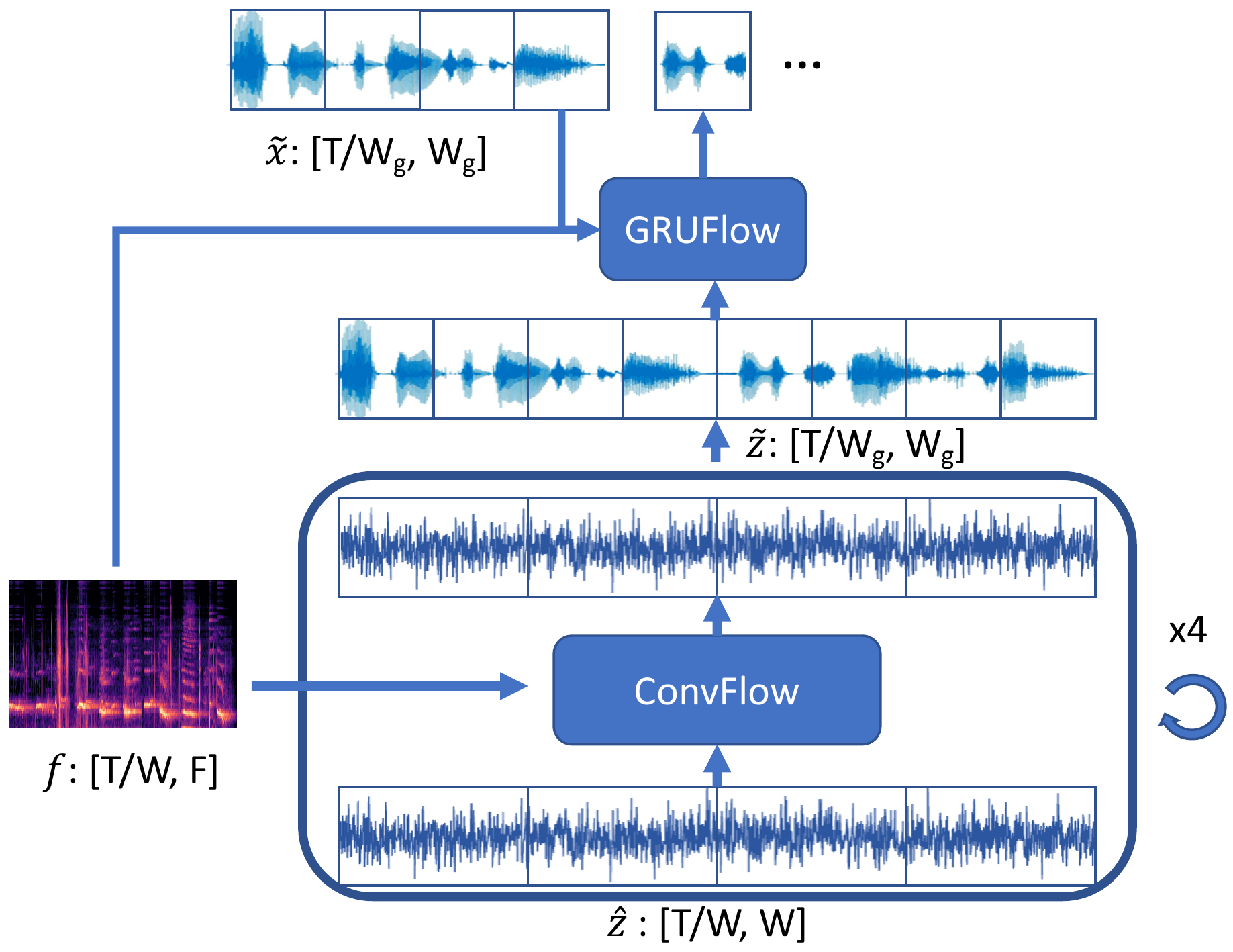}
    \caption{The overall diagram of FBWave. }
    \label{fig:overall}
\end{figure}

\begin{figure}[h]
    \centering
    \includegraphics[width=.9\linewidth]{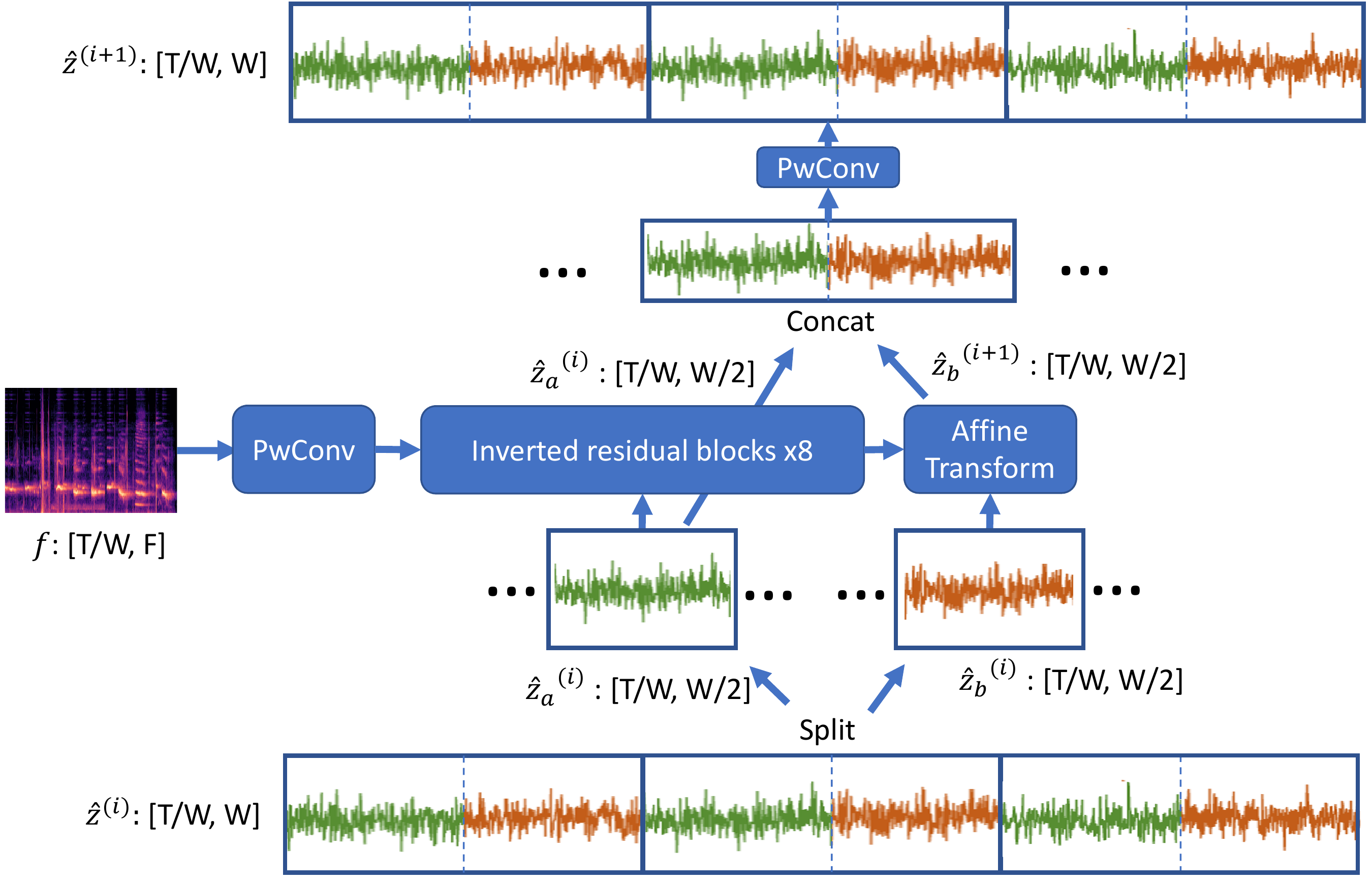}
    \caption{The diagram of ConvFlow. }
    \label{fig:convflow}
\end{figure}

\begin{figure}[h]
    \centering
    \includegraphics[width=\linewidth]{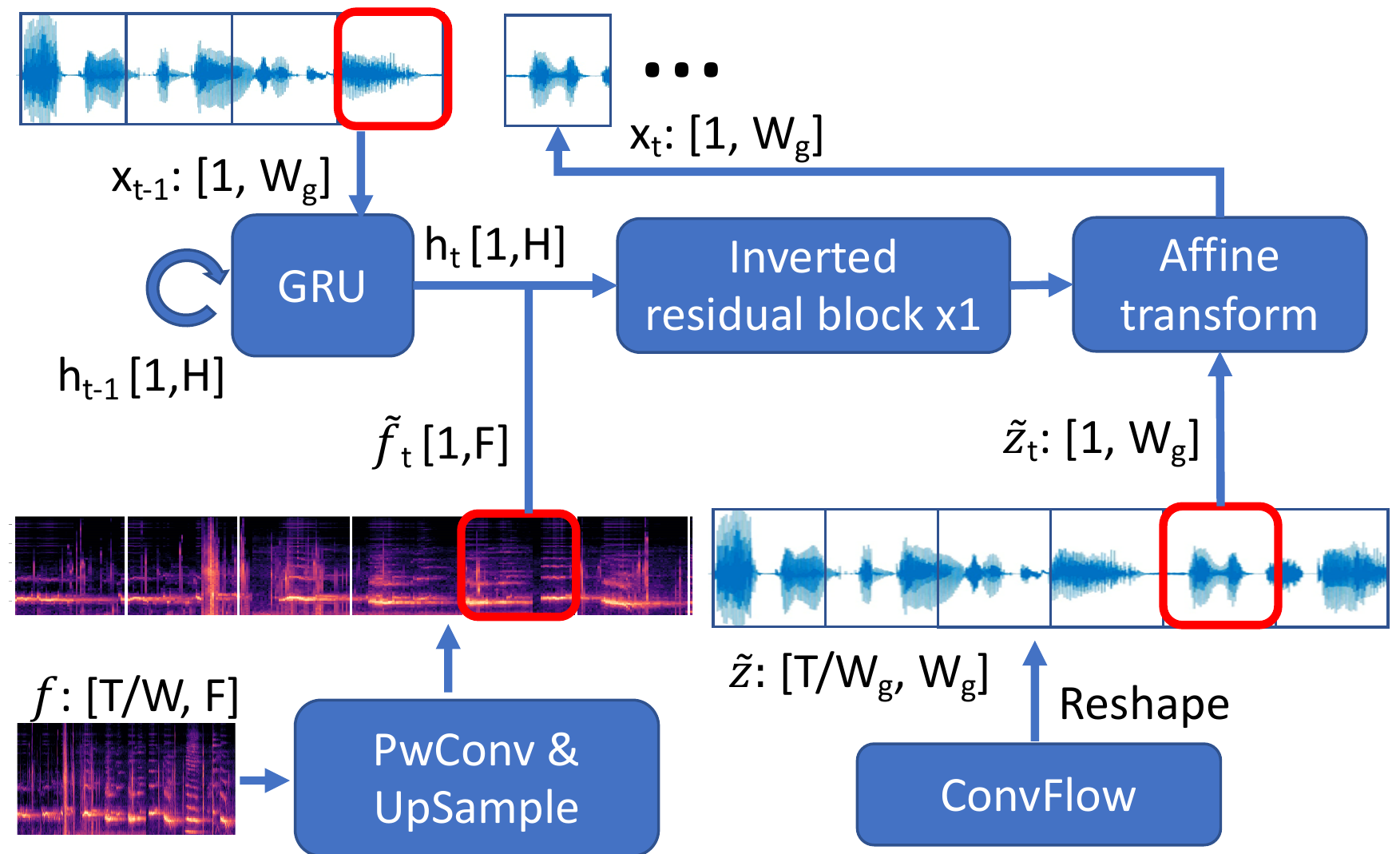}
    \caption{The diagram of GRUFlow. }
    \label{fig:gruflow}
\end{figure}

FBWave is a hybrid flow-based generative model, as shown in Fig. \ref{fig:overall}. It contains a series of invertible transformation functions to convert a simple distribution, such as spherical Gaussian, to a complex audio distribution conditioned on the acoustic features. A generative flow can be described as
\begin{equation}
    \textbf{x} = g_k \circ g_{k-1} \cdot \cdots g_1(\textbf{z}),
    \label{eqn:flow}
\end{equation}
where $\textbf{x} \in \mathbf{R}^{T}$ is the audio sequence with $T$ samples,  $\textbf{z} \in \mathbf{R}^{T}$ is sampled from a simple distribution. $g_i(\cdot)$ can take any form of invertible functions. We adopt two types of functions, a convolution-based non-autoregressive flow function (ConvFlow), as shown in Fig. \ref{fig:convflow}, and a GRU-based autoregressive flow function (GRUFlow), as shown in Fig. \ref{fig:gruflow}. 

\subsection{ConvFlow}

A ConvFlow is similar to the generative flows in Glow\cite{kingma2018glow}, WaveGlow\cite{waveglow}, and Squeezewave\cite{zhai2020squeezewave}. For an input sequence $\mathbf{z}^{(i)} \in \mathbf{R}^T$, a ConvFlow first reshapes $\mathbf{z}^{(i)}$ to $\hat{\mathbf{z}}^{(i)} \in \mathbf{R}^{T/W \times W}$, a sequence of $T/W$ windows with each window containing $W$ contiguous samples. To construct an invertible function, we split $\hat{\mathbf{z}}^{(i)}$ along the second (channel) dimension to two tensors as $\hat{\mathbf{z}}_a^{(i)}, \hat{\mathbf{z}}_b^{(i)} \in \mathbf{R}^{T/W \times W/2}$, and compute 
\begin{equation}
\begin{gathered}
    \mathbf{s}^{(i)}, \mathbf{b}^{(i)} = \text{IRB}(\text{concat}(\hat{\mathbf{z}}_a^{(i)}, \mathbf{f})), \\
    \hat{\mathbf{z}}_b^{(i+1)} = (\hat{\mathbf{z}}_b^{(i)} - \mathbf{b}^{(i)}) / \mathbf{s}^{(i)} , \\
    \hat{\mathbf{z}}^{(i+1)} = \text{PwConv}(\text{concat}(\hat{\mathbf{z}}_a^{(i)}, \hat{\mathbf{z}}_b^{(i+1)})).
\end{gathered}
\label{eqn:convflow}
\end{equation}
$\mathbf{f}\in \mathbf{R}^{T/W \times F}$ is the feature computed by the upstream TTS pipeline. $\text{IRB}(\cdot)$ is a series of inverted residual block functions used to compute affine transformation coefficients $\mathbf{s}^{(i)}, \mathbf{b}^{(i)} \in \mathbf{R}^{T/W \times W/2}$. $\text{PwConv}(\cdot)$ is a 1D point-wise convolution. It computes the output as $\mathbf{\hat{z}}_t^{(i+1)} = \mathbf{K} \mathbf{\hat{z}}_t^{(i)}$ for $t=1, \cdots, T/W$, and $\mathbf{K} \in \mathbf{R}^{W\times W}$ is the weight matrix. 

\begin{figure}[h]
    \centering
    \includegraphics[width=.9\linewidth]{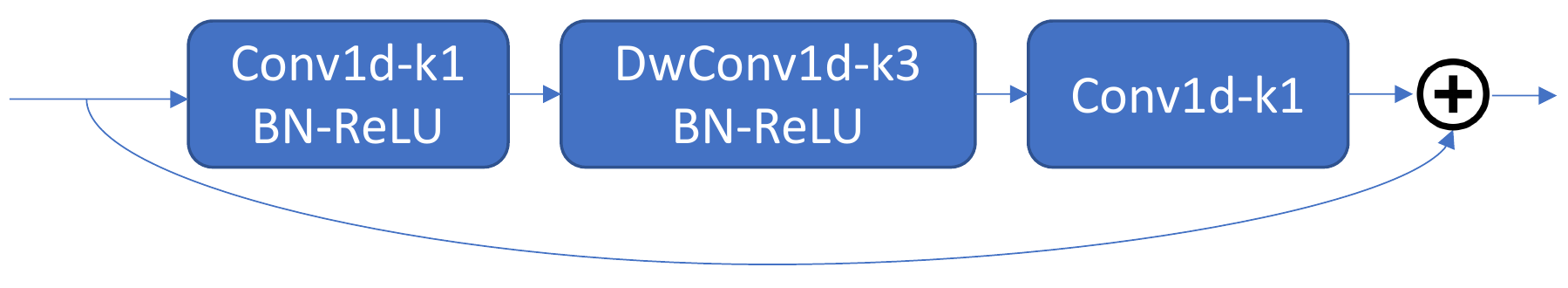}
    \caption{The diagram of an inverted residual block.}
    \label{fig:irb}
\end{figure}

A major difference between FBWave and either WaveGlow or SqueezeWave is the adoption of 1D inverted residual blocks (IRBs) in the ConvFlow. IRBs have become the \textit{de facto} building block for efficient computer vision models \cite{wu2019fbnet,sandler2018mobilenetv2}. As shown in Fig. \ref{fig:irb}, an IRB consists of a point-wise convolution, a depthwise convolution with a kernel size of 3, and another point-wise convolution, in parallel with a skip-connection that adds the input to the output. The input and output have the same channel size $C$, while the intermediate tensor can have a large channel size $E\times C$. The MACs of an inverted residual block is proportional to $\mathcal{O}( EC^2 T / W)$. IRBs are simpler than the WaveNet-like functions used in WaveGlow and its training converges faster. 

\begin{figure}[h]
    \centering
    \includegraphics[width=\linewidth]{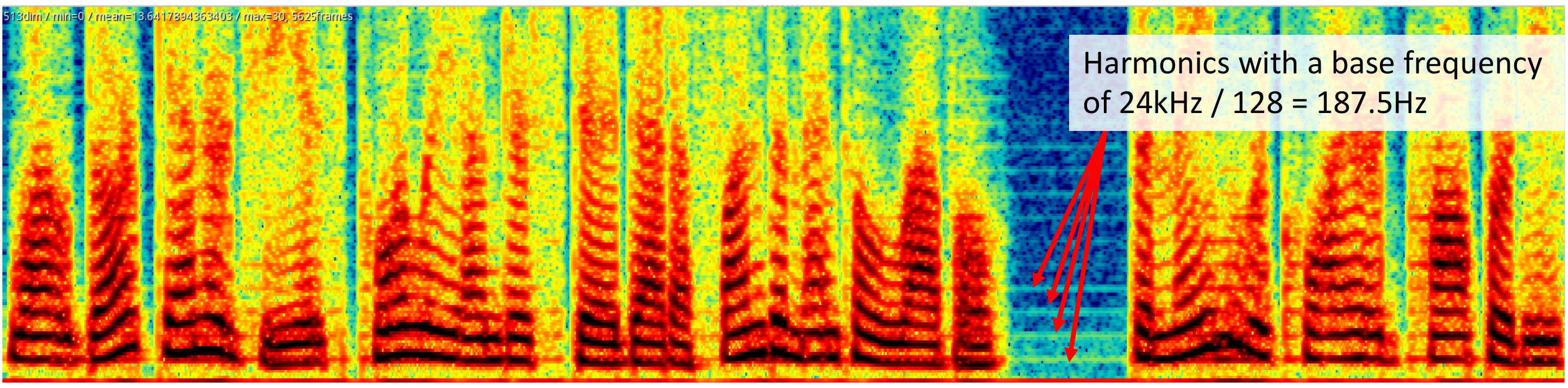}
    \caption{Harmonics observed in the spectrogram of audios generated by SqueezeWave. The base frequency of the harmonics is 24kHz / 128 = 187.5Hz, with 24kHz as the audio frequency and 128 as the window size of SqueezeWave.}
    \label{fig:harmonics}
\end{figure}

ConvFlow functions are non-autoregressive -- samples in different windows are computed without depending on each other. This causes discontinuity between samples in different windows, which leads to notable artifacts in the generated audio. Using SqueezeWave as an example, we analyzed the spectrogram of a snippet of 24kHz audios generated by SqueezeWave. As shown in Fig. \ref{fig:harmonics}, we can observe harmonics with a base frequency of 187.5Hz = 24kHz / 128, where 128 is the window size used by SqueezeWave. Similar artifacts can also be observed in audios generated by WaveGlow. We  observed further quality drop when using such models in a streaming pipeline, where the models are only exposed to a small segment of input features, instead of all the features. 

\subsection{GRUFlow}
To address the limitation of continuity and streamability of ConvFlow, we introduce a GRU-augmented autoregressive flow to process the output ConvFlow functions. At each time step, we use a GRU to take the previous output $\mathbf{\tilde{x}}_{t-1} \in \mathbf{R}^{W_g}$ as input and compute a representation $\mathbf{h}_{t} \in \mathbf{R}^{H}$ encoding the previous states. $W_g < W$ is the window size of the GRUFlow, $H$ is the GRU's state dimension. We concatenate $\mathbf{h}_{t}$ and the upsampled feature $\mathbf{\tilde{f}}_t$ and feed to an IRB to compute the affine transformation coefficients and convert the input as
\begin{equation}
    \begin{gathered}
        \mathbf{h}_t = GRU(\mathbf{\tilde{x}}_{t-1}, \mathbf{h}_{t-1}) \\
        \mathbf{s}_t, \mathbf{b}_t =
        IRB(\text{concat}(\mathbf{h}_t, \mathbf{\tilde{f}}_t)) \\
        \mathbf{\tilde{x}}_t =  (\tilde{\mathbf{z}}_t - \mathbf{b}_t) / \mathbf{s}_t,
    \end{gathered}
    \label{eqn:gruflow}
\end{equation}
GRUFlow achieves better continuity than ConvFlow since $\mathbf{\tilde{x}}_{t}$ is conditioned on previous outputs. This also allows it to be used in a streaming pipeline. We set the IRB's channel size the same as the GRU's hidden state dimension $H$, the MACs for the GRUFlow is proportional to $\mathcal{O}(H^2T/W_g)$. 

\subsection{Training FBWave}
To train FBWave, we maximize the log-likelihood of observed audio sequences $\mathbf{x}$. To compute log-likelihood, we first compute $\mathbf{z}$ from $\mathbf{x}$ by inverting Equation (\ref{eqn:flow}), with each $g_i(\cdot)$ instantiated as Equation (\ref{eqn:convflow}) or (\ref{eqn:gruflow}).  Equation (\ref{eqn:gruflow}) is invertible since given the output $\mathbf{\tilde{x}}_{t-1}$, we can compute $\mathbf{s}_t, \mathbf{b}_t$ and compute the input at step-$t$ as $\mathbf{\hat{z}}_t = \mathbf{\hat{s}}_t \odot \mathbf{\hat{x}}_t + \mathbf{\hat{b}}_t$. To invert Equation (\ref{eqn:convflow}), we first invert the point-wise convolution by $\mathbf{\hat{z}}_t^{(i)} = \mathbf{K}^{-1} \mathbf{\hat{z}}_t^{(i+1)}$, where $\mathbf{K}^{-1}$ is the inverse matrix of $\mathbf{K}$. Next, the affine transformation in (\ref{eqn:convflow}) can be inverted by feeding in half of the output $\hat{\mathbf{z}}_a^{(i+1)}$, which is the same as $\hat{\mathbf{z}}_a^{(i)}$, to compute $\mathbf{s}^{(i)}, \mathbf{b}^{(i)}$, then compute the other half of the input $\hat{\mathbf{z}}_b^{(i)} = \hat{\mathbf{z}}_b^{(i+1)} \odot \mathbf{s}^{(i)} + \mathbf{b}^{(i)}$. 

Next, we compute the log-likelihood
\begin{equation}
\log P(\mathbf{x}) = \log P(\mathbf{z}) + \sum_{i=1}^k \log|\det(Jg_i^{-1}(\mathbf{z}^{(i+1)})|,
\label{eqn:loss}
\end{equation}
where $\log \deg(Jg_i^{-1}(\mathbf{z}^{(i+1)}))$ denotes the log determinant of the Jacobian of $g_i^{-1}(\mathbf{z}^{(i+1)})$. For affine transformations in GRUFlow and ConvFlow, it can be computed as
\begin{equation}
    \log |\deg(Jg_{\text{affine}}^{-1}(\mathbf{z}^{(i+1)}))| = \sum_{t,k} | \mathbf{s}_{t,k}|.
\end{equation}
For ConvFlow, we also need to compute
\begin{equation}
    \log |\deg(Jg_{\text{PwConv}}^{-1}(\mathbf{z}^{(i+1)}))| = T/W \times  \log |\det (K^{-1})|. 
\end{equation}
Finally, to compute $\log P(\mathbf{z})$ in Equation (\ref{eqn:loss}), previous work \cite{waveglow,zhai2020squeezewave,kingma2018glow} assume that $\mathbf{z}$ comes from a spherical Gaussian distribution and compute it as $\log P(\mathbf{z}) = - \sum_{ij} \mathbf{z}^2_{ij} / (2\sigma^2)$. In our experiment, we found that using Laplace distribution $P(\mathbf{z})$ leads to more stable training and better audio quality. So instead we compute   $\log P(\mathbf{z}) = - \sum_{ij} |\mathbf{z}_{ij}| / \sigma$.

\subsection{Scalable architectures}
The computational cost of FBWave can be controlled by a few architecture hyperparameters. For a given sequence length of $T$, the MACs of ConvFlow is proportional to $\mathcal{O}(C^2ET/W)$, where $C$ is the channel size of the IRB, $E$ is the expansion ratio, and $W$ is the window size. For the GRUFlow, its MACs is proportional to $\mathcal{O}(H^2T/W_g)$, with $H$ as GRU's hidden dimension and the IRB's channel size, and $W_g$ as its window size. By tunning $C, H, E, W_g, W$, we obtain a wide range of models to fit in the compute budgets of different devices. 

\section{Experiments}
We conduct experiments to evaluate the audio quality and the efficiency of FBWave. In many previous works \cite{waveglow,zhai2020squeezewave}, a vocoder's performance is evaluated by synthesizing audios from mel-spectrograms extracted from ground-truth audios. However, generating audios from grouth-truth acoustic features is a much simpler task than synthesizing using features predicted by the upstream TTS models.  Using the ground truth acoustic features, even parametric vocoders such as the STRAIGHT vocoder \cite{kawahara2006straight} or the WORLD vocoder \cite{morise2016world} can generate high quality audios. So instead, we evaluate FBWave by using it in an industrial TTS pipeline.

Our TTS pipeline (see Section 4.2 of \cite{gao2020interactive} for details) consists of a prosody model that predicts the durations of each phone and an acoustic model that predicts a 13-dimensional Mel-frequency cepstral coefficients (MFCCs) feature, a 1-dimensional fundamental frequency (log-F0) feature, and a 5-dimensional periodicity feature for each frame. Our feature frame has a window size of $10.666$ms and frame shift of $5.333$ms. With an audio sampling rate at 24kHz, the spectral feature's frequency is 128 times slower than the audio frequency. We train FBWave using a dataset recorded in a voice production studio by a professional voice talent. It contains 40,244 utterance, approximately 40 hours of training data.
 
We train FBWave using the Adam optimizer \cite{kingma2014adam} with a batch size of 64, a learning rate of 1e-3, and we decay the learning rate by $\exp$(-5e-3) per epoch. Each audio segment in a batch consists of 8192 samples. We use batch normalization \cite{ioffe2015batch} as it leads to faster convergence than weight normalization \cite{salimans2016weight} used by \cite{waveglow,zhai2020squeezewave,kumar2019melgan}. Our model is extremely efficient to train so even the largest model can be trained using \textbf{a single} Nvidia V100 GPU. In comparison, training WaveGlow with a batch size of 24 requires \textbf{8 V100 GPUs}.

To evaluate the efficiency, we measure the number of MACs (multiplication-and-accumulate operations) needed to generate 1 second of 24 kHz audio. The MAC count is hardware-agnostic, which is ideal for evaluating models targeting different devices. To evaluate the audio quality, we conduct mean opinion score (MOS) tests on the generated audios with 400 participants who rate a each sample with a score between 1-5 (1:bad - 5:excellent). The test set includes 60 utterances ranging from 1 to over 30 seconds. 

First, we compare FBWave with baseline models, including WaveRNN \cite{WaveRNN},
SqueezeWave \cite{zhai2020squeezewave}, and a classic source-filter parametric model similar to \cite{kawahara2006straight}. We report the MOS score and the MAC count in Table \ref{tab:result-with-baseline}. As we can see, FBWave achieves competitive MOS scores compared to WaveRNN and WaveRNN (94\% sparsity), while the computational cost is 40x and 3.8x smaller, respectively. Compared with the parametric model and SqueezeWave\cite{zhai2020squeezewave}, FBWave models achieve much higher MOS scores. After quantization, FBWave can achieve an streaming inference speed of 0.7 RTF (real-time factor) on a Samsung S8 smart phone. A qualitative evaluation of the audios also reveal that the quality of FBWave is similar to WaveRNN models and much better than the parametric model and SqueezeWave. Audios generated by different models for the MOS test are available at \url{https://bichenwu09.github.io/vocoder_demos}.

\begin{figure}[h]
    \centering
    \includegraphics[width=.8\linewidth]{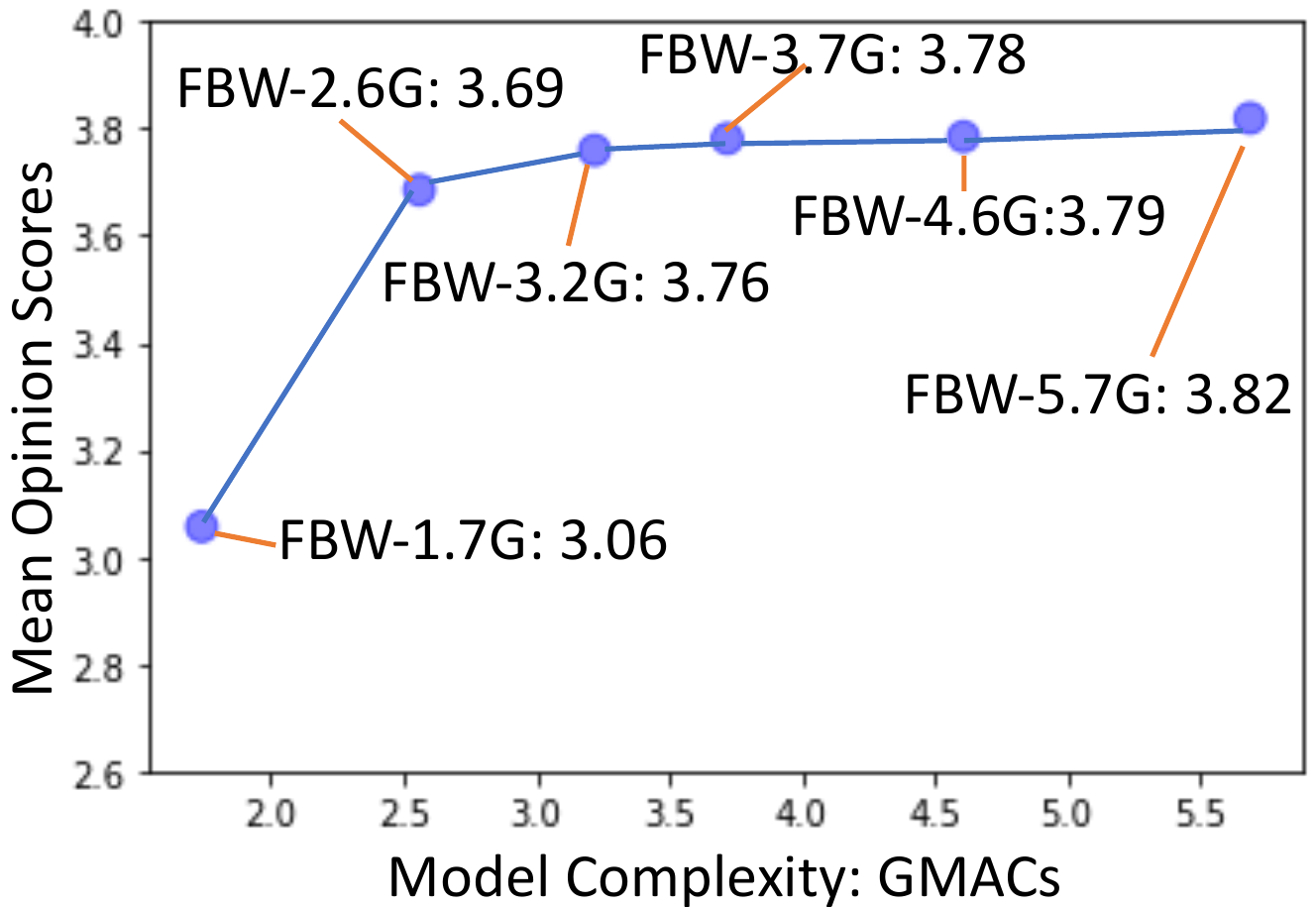}
    \caption{Audio quality (MOS scores) and the efficiency (MACS) trade-off of FBWave family models. }
    \label{fig:mos-vs-macs}
    \vspace{-10pt}
\end{figure}

By tuning the architecture hyperparameters of FBWave, we obtain a family of efficient models. We train them and run another MOS test to compare these models, and plot their MOS scores and MACs in Fig. \ref{fig:mos-vs-macs}. As we can see, as we shrink FBWave models, the MOS scores drop very slowly. Even our smallest model, FBW-1.7G, can deliver acceptable audio quality while its computational cost is 109x smaller than WaveRNN. This family of FBWave models can be adapted to different target devices with diverse computational capacity.  

\begin{table}[]
\centering
\begin{tabular}{l|cc}
\hline
                   & MOS             & GMACs \\ \hline
Ground truth       & 4.41 $\pm$ 0.33 & -     \\ \hline
WaveRNN \cite{WaveRNN}           & 4.13 $\pm$ 0.04 & 184.6 \\
WaveRNN (sparse) \cite{WaveRNN}  & 4.09 $\pm$ 0.04 & 17.5  \\
Parametric Model \cite{kawahara2006straight} & 3.08 $\pm$ 0.06 & -     \\
SqueezeWave \cite{zhai2020squeezewave}       & 2.70 $\pm$ 0.06 & 3.8   \\ 
FBWave-4.6G (ours) & 3.81 $\pm$ 0.04 & 4.6   \\
\hline
\end{tabular}
\caption{Comparing FBWave with baseline models. 
}
\label{tab:result-with-baseline}
\vspace{-10pt}
\end{table}

\section{Conclusions}
We present FBWave, a family of efficient and scalable neural vocoders for edge-based TTS. We propose a hybrid-flow architecture that uses an autoregressive flow on top of a parallel flow such that the model can be used in a streaming pipeline for on-device TTS. Moreover, the autoregressive flow mitigates the discontinuity in audios generated by parallel flows, which leads to significant quality improvement, as confirmed by MOS tests and qualitative evaluations. By controlling the model using architecture parameters, we obtain a family of FBWave models that are 40 - 109x more efficient than a WaveRNN model while still deliver competitive audio quality under different constraints for different target devices. 

\newpage

\bibliographystyle{IEEEbib}
\bibliography{refs}

\end{document}